\documentclass[prl,aps]{revtex4}
\usepackage{graphicx,latexsym}

\begin{document}
\title{Instability of charged and rotating naked singularities}

\author{Gustavo Dotti$^1$, Reinaldo J. Gleiser$^1$, Jorge Pullin$^2$}
\affiliation { 1. Facultad de Matem\'atica, Astronom\'{\i}a y
F\'{\i}sica, Universidad Nacional de C\'ordoba, Ciudad
Universitatis, (5000) C\'ordoba,\ Argentina\\ 2. Department of
Physics and Astronomy, Louisiana State University, Baton Rouge, LA
70803-4001}

\begin{abstract}
  We provide evidence that ``super-extremal'' black hole space-times
  (either with charge larger than mass or angular momentum larger than
  mass), which contain naked singularities, are unstable under
  linearized perturbations.  This is given by an infinite family of
  exact unstable solutions in the charged non rotating case, and by a
  set of (unstable) numerical solutions in the rotating case.  These
  results may be relevant to the expectation that these space-times
  cannot be the endpoint of physical gravitational collapse.
\end{abstract}

\maketitle It is well known that the equations of general
relativity admit exact solutions that contain singularities. In
some cases, as in black hole space-times, the singularities are
``hidden'' behind horizons, that is, regions that cannot
communicate causally with the rest of space-time.  However, the
same solutions that describe black holes can, for certain choices
of their parameters, describe ``naked'' singularities. This is the
case of a charged (Reissner--Nordstr\"om) space-time with $|Q|>M$
or a rotating (Kerr) solution with angular momentum $a>M$. These
represent problematic space-times, since the singularities can
communicate causally with the exterior. There has always been the
expectation that these solutions are not the endpoint of
physically relevant collapsing matter. This whole issue frames
itself in the context of the ``cosmic censorship'' hypothesis of
Penrose that loosely stated claims no naked singularities are
formed by gravitational collapse. Several arguments have been put
forward that show that one indeed cannot drop charges into a
charged black hole to turn it into ``super-extremality'' or spin
up a Kerr black hole into a similar regime \cite{wald}.

In this paper we would like to argue that ``super-extremal''
space-times are unstable under linearized perturbations. This would
further strengthen the idea that they cannot be the endpoint of
gravitational collapse. We start our discussion by considering a
spherically symmetric, charged space-time described by the
Reissner--Nordstr\"om solution of the Einstein equations,
\begin{equation}
\label{Schw1}
ds^2 = -f dt^2 +
(1/f) dr^2 +r^2 d\Omega^2
\end{equation}
where $f=(r^2-2rM+Q^2)/r^2$, and $d \Omega^2$ is the standard line
element on the unit sphere. The real (positive or negative)
constants $M$ and $Q$ can be respectively identified with the
Newtonian mass and the total (electric) charge of the source. We
consider only the case $M > 0$ in what follows. The case $M<0$,
$Q=0$ has been previusly analyzed in \cite{ghi} and
\cite{dottigleiser}.

The range of $r$ in (\ref{Schw1}) is limited by the singularities of
the metric coefficients. These occur for $r=0$ and $r_{\pm} = M \pm
\sqrt{M^2-Q^2}$.  For $M>|Q|$ the singularity at $r=r_{+}$
corresponds to a regular horizon, and the solution may be extended
to $r<r_{+}$. The resulting space-time is known as a ``charged black
hole''. The horizon hides a curvature singularity for $r=0$,
rendering the portion of space time outside the horizon globally
hyperbolic. However, if $|Q|>M$, there are no horizons, and the
space-time contains a ``naked" singularity for $r=0$.

The question of the (linear) stability of both charged and uncharged
black hole space-times has been analyzed by many authors, starting
with the papers by Regge and Wheeler, and Zerilli
\cite{reggewheeler} (see  \cite{kodamaishibashi} for a recent
formulation in arbitrary dimensions). To simplify the discussion we
shall restrict to polar (scalar) metric perturbations in  the
Regge--Wheeler gauge \cite{reggewheeler,kodamaishibashi,prs}. The
metric and electromagnetic scalar perturbations for the angular
modes with fixed $\ell$, ($\ell=2,3,...$),  can then be encoded in two
functions, $\Phi_{i}(r,t), i=1,2,$ that satisfy the equations
\begin{equation}
\label{eqPhip1}
   \frac{\partial \Phi_ i}{\partial t^2} = \frac{\partial \Phi_ i}
{\partial r*^2} - \left(
 \beta_{i} \frac{d f_i}{d r*} + \beta_i{}^2 f_i{}^2 + \kappa f_i
 \right) \Phi_i \equiv -{\cal A}_i \Phi_i,
\end{equation}
where $\kappa = (\ell-1) \ell (\ell+1) (\ell+2)$, $\;\;\beta_i = 3
(M + (-1)^{i} \mu)$, and,
\begin{equation}
f_i = \frac{f}{\beta_i + (\ell -1)(\ell+2) r^2}
\end{equation}
with $\mu=\sqrt{M^2+4 Q^2 (\ell-1)(\ell+2)/9}$, and $r^*$ the
``tortoise" coordinate defined by $dr^*/dr = 1/f$,
\begin{equation}
r^* = r + M \ln \left( \frac{r^2-2Mr+Q^2}{M^2} \right) +
\frac{2M^2-Q^2}{\sqrt{Q^2-M^2}} \arctan \left(
\frac{r-M}{\sqrt{Q^2-M^2}} \right) + c.
\end{equation}
These have been analyzed in the case $M^2 > Q^2$, where there is a
regular horizon, hiding the curvature singularity at $r=0$. On the
other hand, we are here interested in the case $Q^2 > M^2$, with
$M \geq 0$. In this case, the range of $r$ is the full interval
$(0, +\infty)$, so it is crucial to establish what boundary
conditions are acceptable for $r \rightarrow 0$. If we assume that
the solutions are regular near $r=0$, and expand $\Phi_{i}$ in
power series in $r$, with time dependent coefficients, replacing
these expansions in (\ref{eqPhip1}), we find,
\begin{equation}
{\Phi_{i}} \left( t,r \right) = b _{{i}} \left( t \right)
r+c_{{i}}
 \left( t \right) {r}^{2}+ \left(
 \frac {[2 (\ell+2)(\ell-1)Q^2+3 M (M -(-1)^i\mu)] b_i(t)} {8 Q^4}
+{\frac {M c_{{i}} \left( t \right) } {{Q}^{2}}} \right)
{r}^{3}+{\cal{O}}( {r}^{4})
\end{equation}
where $b_{1,2}$, $c_{1,2}$ are, in principle, arbitrary functions
of $t$. Notice that both $\Phi_{i}$ contain two arbitrary
functions of $t$ and, therefore, these expansions represent the
general behavior of these functions near $r=0$.  We may replace
the expansions in the expressions for the perturbations of the
metric in the Regge-Wheeler gauge near $r=0$. It is then
straightforward to obtain, for instance, the Kretschmann
invariant, ${\cal{K}}  = R_{abcd} R^{abcd}$ up to first order in
the perturbations. We recall that for the background metric we
have $ {\cal{K}} = {56 Q^4}/{r^8} +{\cal{O}}(r^{-7})$. The
explicit expression including first order terms is rather long to
give explicitly here. After replacing the metric coefficients in
${\cal{K}}$ and some simplification we find that, in general, the
perturbations introduce singular contributions that diverge faster
than the background, unless the following conditions are imposed
\begin{equation}
\label{boundary1}
c_{{1}}( t) =  {\frac{ 3 \left( M +\mu \right)
}{4{Q}^{2}}} b_{{1}}( t) \qquad c_{{2}}( t )  =  {\frac{ 3 \left(
M -\mu \right) }{4{Q}^{2}}}   b_{{2}}( t )
\end{equation}
in which case the Kretschmann invariant has the behavior given
above. We shall take (\ref{boundary1}) as the appropriate boundary
condition for the perturbations in what follows. We therefore
exclude perturbations that might either modify the nature of the
singularity, or take us beyond the domain of linear perturbations.
We remark that these are the most restrictive boundary conditions
that can be imposed at the singularity without making the solution
trivial. We will see, nevertheless, that even under this
restrictions the perturbation equations have unstable solutions.

The usual procedure for analyzing stability is to consider solutions
of (\ref{eqPhip1}) of the form $\Phi_{i}(t,r)= \exp(-i \omega t)
\phi_{i}(r)$ and look for pure imaginary values of $\omega$.
Replacement of this Ansatz in (\ref{eqPhip1}) leads to second order
ordinary differential equations for $\phi_{i}$. For the negative
mass Schwarzschild singularity, analytic expressions for unstable
modes of arbitrary $\ell$, satisfying appropriate boundary
conditions, were obtained in \cite{dottigleiser}. It was then
noticed by Cardoso and Cavaglia \cite{cardoso} that these unstable
modes agree with the {\em algebraically special} (AS) solutions  in
\cite{prs}, which, although irrelevant as perturbations in the black
hole ($M>0$) regime -due to their behavior at the horizon- satisfy
appropriate boundary conditions in the nakedly singular case. The AS
modes for the Reissner--Nordstr\"om spacetime have pure imaginary
frequency $\omega = -ik$ with $k = -\kappa/(2\beta_1)$ (so that $k >
0$). They can be constructed following \cite{prs}, the result being
\begin{equation} \label{rns}
\Phi_2 = 0 , \hspace{1cm} \Phi_1 = \left[ C_1 \chi(r) + C_2 \chi(r)
 \int^r \frac{du}{\chi(u)^2} \right] e^{kt},
\end{equation}
where
\begin{equation} \label{rns2}
\chi(r) = \frac{r e^{-kr*} }{(\ell-1)(\ell+2)r + \beta_1} ,
\hspace{0.5cm} \mbox{with} \hspace{0.5cm} k = \frac{(\ell-1) \ell
(\ell+1) (\ell+2)}{2\left(\sqrt{9 M^2+4 Q^2 (\ell-1)(\ell+2)}-3
M\right)}
\end{equation}
We can readily check that for all $Q^2 > M^2$, and $\ell$,
unstable solutions satisfying appropriate boundary conditions are
obtained by simply setting $C_2=0$ in (\ref{rns}). Nevertheless,
since $\beta_1 < 0$, the resulting solutions for $\Phi_1$ are
singular for $r=-\beta_1/[(\ell-1)(\ell+2)]$. We notice, however,
that replacing these expressions into those for the metric
perturbations we find that this ``kinematic singularity'' is
absent both in the electromagnetic field and the perturbations of
the metric coefficients, and, therefore, for finite $t$ these
solutions correspond to the evolution of regular perturbations
that can be made initially arbitrarily small as compared to the background
(see \cite{dottigleiser} for a discussion of an analogous
situation in the $Q=0, M < 0$ case). This is confirmed by an
explicit computation of ${\cal{K}}$ to first order in the
perturbations. The result (exact in $r$) is,

\begin{equation}\label{kretgen}
{\cal{K}}  = \frac {8( 6r^2 M^2-12 rM{Q}^{2}+7{Q}^{4})}{{r}^{8}}+
\frac {C {\ell}
 \left( {\ell}+1 \right) \left( {\ell}+2 \right) \left( {\ell}-1 \right)
 \left( Mr-{Q}^{2} \right) {Q}^{2}\,{e^{kt}}  e^{-k r*} Y_{\ell m}
  }{{r}^{7} \left( -M+\mu \right) ^{2}}
\end{equation}
where the first term on the R.H.S of (\ref{kretgen}) corresponds to
the background, and $C$ is an arbitrary constant. We therefore see
that {\em for all $Q^2 > M^2$}, there exist divergent perturbations of
the Reissner--Nordstr\"om spacetime {\em for all $\ell \geq 2$}.
Notice that if $Q^2<M^2$, the $\exp(-kr*)$ factor in (\ref{rns2})
gives a divergent behavior at the horizon for any choice of $C_i$,
since $r* \simeq M \ln (r-r^+)$ if $r \simeq r^+, r>r^+$, and $k > 0$,
and, therefore, (\ref{rns2}) cannot be considered an
acceptable perturbation of the Reissner--Nordstr\"om spacetime.\\
The issue of dynamics in non-globally-hyperbolic static space-times is
a subtle one, some aspects of which were analyzed in a sequel of
papers by Wald \cite{wi1} and Wald and Ishibashi \cite{wi2,wi3}. In
\cite{wi1,wi2} the Klein Gordon equation (with positive mass) is cast
into the form (\ref{eqPhip1}).  A similar form for gravitational
perturbations on AdS backgrounds of arbitrary dimensions is given in
\cite{wi3}. In all cases, the fact that the analogous of the operator
${\cal{A}}_i$ in eq. (\ref{eqPhip1}) is positive definite on an
appropriate function space, allows to globally define the dynamics of
the field (see, e.g., eq.(4) in \cite{wi3}). A straightforward
consequence of the existence of the exponentially growing modes
(\ref{rns})-(\ref{rns2}) is, however, that ${\cal A}_1$ in equation
(\ref{eqPhip1}) is {\em not} positive definite on the function space
of physically relevant perturbations, i.e., those that do not increase
the degree of divergence of the scalar invariants as the singularity
is approached. Thus, gravitational perturbations around super extremal
charged black holes lie outside the scope of \cite{wi1,wi2,wi3}. A
similar situation was found in the non-globally-hyperbolic negative
mass Schwarzschild solution, and is discussed in detail in
\cite{dottigleiser}.

 We now turn our attention to the uncharged, rotating Kerr
space-times with $a>M$. The linearized perturbations were first
studied by Teukolsky \cite{teukolsky}, who showed that they could
be captured in a master equation in terms of a linearized tetrad
component $\psi$ of the Weyl tensor. The equation can be separated
by assuming $\psi=F(r,\theta,\varphi) \exp(-i\omega t)$ and
$F(r,\theta,\varphi)= S^m_\ell(\theta) \exp(im\varphi)
R_{\omega,\ell,m}(r)$, which leads to a coupled system for $S$ and
$R$,
\begin{eqnarray} \label{ta}
{1\over \sin\theta} {d \over d\theta}\left(\sin\theta {d S\over
d\theta}\right)+\left(a^2\omega^2\cos^2\theta-{m^2\over
\sin^2\theta} -2 a \omega s \cos\theta -{2 m s \cos\theta\over
\sin^2\theta} -s^2 \cot^2 \theta +E -s^2\right)S &=& 0 \\
\label{10}
\Delta {d^2 R \over dr^2} +2 (s+1)(r-M){dR\over dr} +\left\{
{K^2-2is(r-M)K\over \Delta}+4ir\omega s -\lambda\right\}R &=&0
\label{11}
\end{eqnarray}
where $s=\pm2$, $\Delta=r^2 -2 Mr +a^2$, and $K=(r^2+a^2)\omega -a
m$. The eigenvalues $E$ are determined by regularity conditions on
$S(\theta)$ for $\theta= 0, \pi$,
 and $\lambda=E-2am\omega+a^2
\omega^2-s(s+1)$.

For non extremal Kerr black holes ($a^2<M^2$), where a horizon is
present, the region outside the horizon is globally hyperbolic, and
one can show stability of the perturbations under appropriate
boundary conditions at the horizon and for $r \to \infty$. However,
for $a^2> M^2$, the space-time is not globally hyperbolic, and even
though (\ref{ta}) still applies, the question of the stability is
certainly more subtle. As a preliminary Ansatz, we will assume that
even in this case any acceptable unstable perturbation (solution of
(\ref{ta}) and (\ref{10})) must fulfill at least the requirements
that i) it can be made arbitrarily small at some chosen time, and
ii) that it grows exponentially in time. We think that prior to the
present analysis, it was not known if even these modest requirements
could in fact be satisfied.

As shown in \cite{prs}, AS modes $\propto \exp(-i \omega t + im
\phi)$ exist for the Teukolsky equation if the Starobinsky constant
vanishes. This condition and the regularity condition
$S(\theta=0,\pi)=0$ required in (\ref{ta}), impose two constraints
on $\lambda$ and $\omega$ that may have complex $\omega$ solutions
\cite{prs}. These solutions are of the form $R(r) = (A + B r + C r^2
+ D r^3) \exp(i \omega r*)$ \cite{prs}, where $r* \to r$ as $|r| \to
\infty$, and, since the domain of interest when $a^2> M^2$ is
$-\infty < r < \infty$, AS modes always diverge in one of the
asymptotic regions. Therefore, we do not consider AS modes as
relevant perturbations of the super-extremal Kerr solution because
they do not satisfy our Ansatz.

In general, solutions of the Teukolsky  system can be carried out
numerically, as first discussed by Press and Teukolsky
\cite{pressteukolsky}. In our case we started by assuming that
$\omega=ik$ with $k>0$ to seek for solutions that grow with time.
For simplicity we set $m=0$, and chose $E$ to be real, so the
equation for $R$ had real coefficients. The equation for $S$ is
complex.  We integrated (\ref{10}) using a power series in
$x=\cos\theta$ around $x=0$, up to the highest order allowed by
the particular computer implementation that we used, which in our case was
$x^{27}$, but we checked that essentially the same results were
already obtained using expansions up to order 20 or higher.
Regularity of S for $\theta= 0,\pi$, implies $S(x=\pm1)=0$.
Imposing this condition yields a lengthy yet polynomial
relationship between $E$ and $k$. We chose the lowest real value
of $E$ given $k$. With this value of $E$ we solved numerically
equation (\ref{11}) using a shooting method. To set up the
shooting method we worked out asymptotic approximations to the
solution for large values of $|r|$ (in the super-extremal case
Kerr can be extended through the ``ring'' singularity to negative
values of $r$). Indicating with $\pm$ the cases $r \to \pm
\infty$, they are of the form,
\begin{equation}
R(r)^\pm = e^{-k|r| -(c^{\pm}\pm k M )\ln(r^2)}\left[b^\pm_0+
{b^\pm_1 \over r}+{b^\pm_2 \over r^2}+\cdots\right]
\end{equation}
where $c^{-}=1/2$, $c^+= s+1/2$, and the constants $b^{\pm}$ are
adjusted so that (\ref{11}) is satisfied to the given order. These
expansions were used to generate numerical initial data for the
shooting algorithm, typically around $r=\pm 10$ ($M=1$). For $a$
we took values in the range $1.1-1.4$.  The value of $k$ was then
varied until the algorithm yielded a finite solution. A typical
solution for $s=-2$ is shown in figure 1. Similar results, for the
same values of $k$, were obtained for $s=+2$, (i.e., for $\Psi_0$)
as expected from the non vanishing of the Starobinsky constant
\cite{prs}.
\begin{figure}
\centerline{\includegraphics[height=8cm]{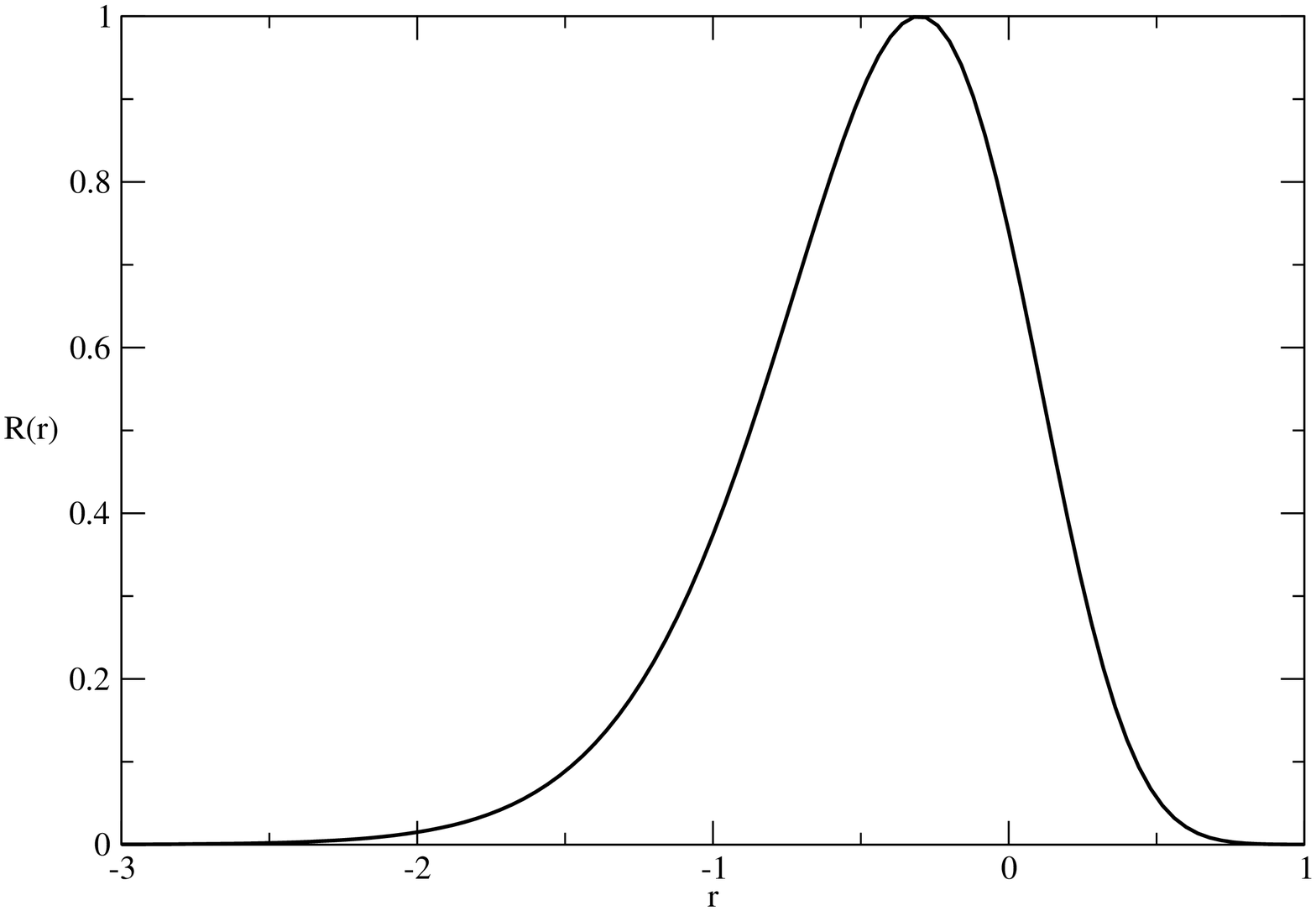}
\includegraphics[height=8cm]{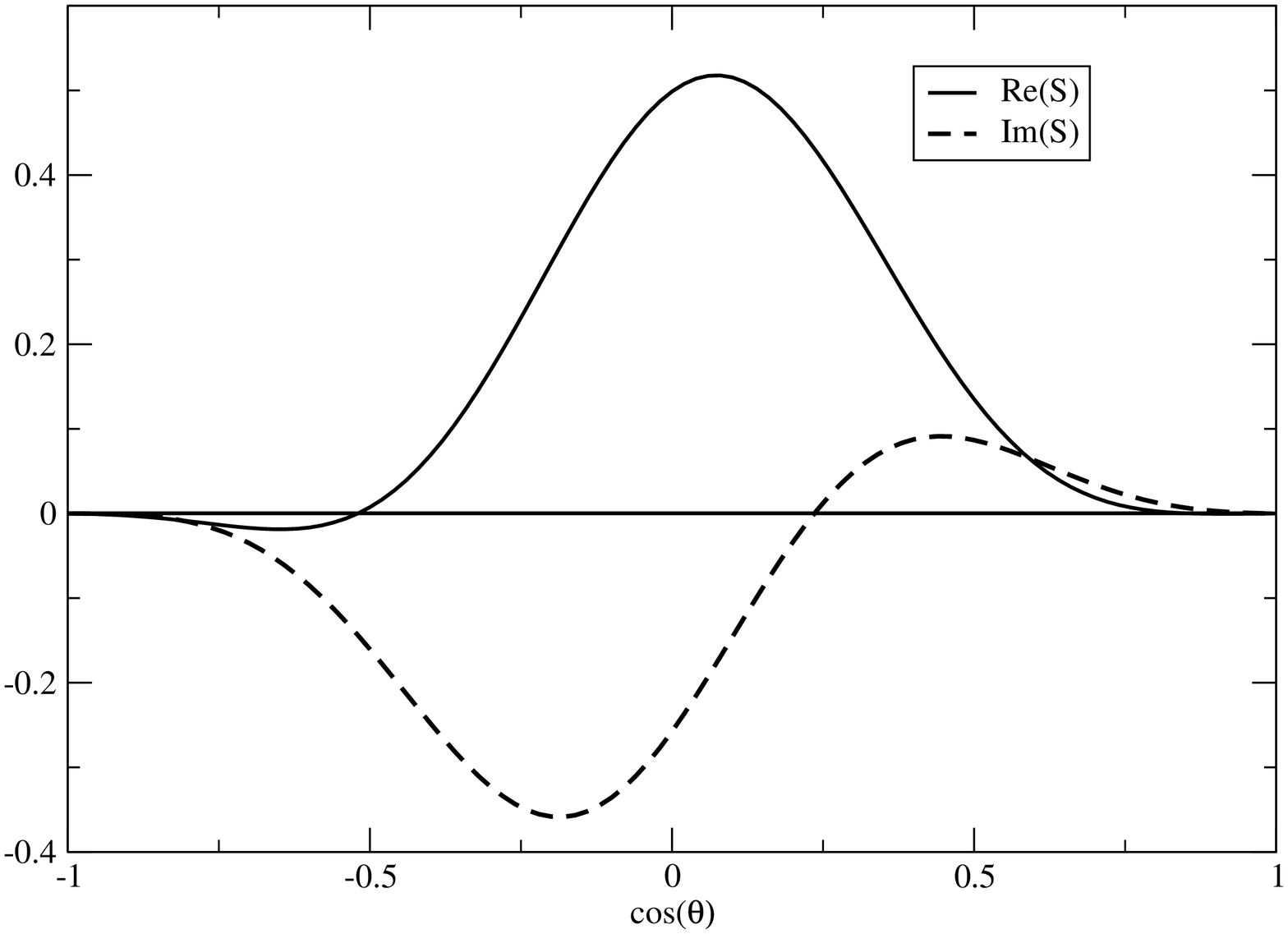}}
\vspace{-1cm} \caption{The left panel shows a typical solution for
the radial part of the perturbation $R(r)$ in the Kerr case,
obtained for $s=-2$, $M=1$, and $a=1.4$. The amplitude is arbitrary
since this is a linear problem, and has been normalized so that
$R(r)=1$ at its maximum. The right panel shows the corresponding
real and imaginary parts of the angular function $S(\theta)$. The
integration gave $k=7.07...$, and $E=17.37...$}
\end{figure}

The shooting algorithm was implemented in Maple using the built in
Runge--Kutta integrator. We checked that the solution was
insensitive to the error tolerance of the integrator and to the
departure point of the shooting. The solution has an exponential
behavior that limits in practice how far in the range of $r$ can
we start the shooting procedure before running into machine
precision problems. We verified the solution by shooting both from
large negative and positive values of $r$. The conclusion of the
numerical analysis is that solutions with $k>0$ exist, at least
for the values of the other parameters chosen, and seem to be
numerically robust. Further studies are needed to confirm the
ranges of values of the various parameters for which unstable
solutions exist.

It should be stressed that any solution $\psi$ of the $s=-2$
Teukolsky equation is a ``Debye potential'' from which a metric
perturbation $h_{ab}$ that solves the linearized Einstein equations
around the Kerr background can be constructed.  $h_{ab}$ is obtained
by applying a second order linear differential operator to $\psi$
(see \cite{chirinovski}).  Using the solution above as a Debye
potential gives a linearized solution $h_{ab} \propto \exp(kt)$,
and, therefore, unstable, of Einstein's equations around a
super-extremal Kerr space-time. However, the explicit expression of
the metric perturbation is so lengthy  that it is hard to use it in
further computations  to evaluate, e.g.,  the perturbation effects
at the ring singularity. One may want to compute the perturbed
values of the Riemann invariants. A basis of algebraic invariants
for the Riemann tensor in vacuum is given by the complex scalar
fields \cite{carminati}
\begin{eqnarray} \nonumber
W_1 &:=& \frac{1}{4} \hat C_{abcd} \hat C ^{abcd} = 2
\Psi_0 \Psi_4 - 8\Psi_1 \Psi_3 + 6 (\Psi_2)^2\\
 W_2 &:=& -\frac{1}{8} \hat C_{abcd} \hat C ^{cd}{}_{ef} \hat
 C^{efab} = 6 \Psi_4 \Psi_0 \Psi_2 - 6 (\Psi_2)^3 - 6
 (\Psi_1)^2 \Psi_4 - 6 (\Psi_3)^2\Psi_0 + 12 \Psi_2 \Psi_1 \Psi_3
 \label{invs}
\end{eqnarray}
where $C_{abcd}$ the Weyl tensor and $\hat C_{abcd}:= \left(
C_{abcd} + i {}^*C_{abcd} \right)$ (note that the Kretschmann
invariant is given by the real part of $W_1$.) Since the only
nonzero Weyl scalar for the Kerr background is $\Psi_2$
\cite{teukolsky,chandra2}, the linearization of (\ref{invs}) yields
\begin{equation}
\delta W_1 =  12 \Psi_2{} \; \delta \Psi_2 ,  \;\;\; \delta  W_2 =
-18 (\Psi_2)^2 \; \;\delta \Psi_2.
\end{equation}
However, $\delta \Psi_2 = 0$ for arbitrary perturbations (with the
exception of  stationary, axially symmetric perturbations) of the
Kerr spacetime \cite{chandra2}, and thus algebraic invariants are
not modified to first order. A similar situation was found for the
unstable modes of the negative mass Schwarzschild spacetime
\cite{dottigleiser}, for which the perturbation effect on the
singularity was then analyzed by computing {\em differential}
invariants of the Riemann tensor, an approach that is hard to
implement in this case, in view of the above mentioned length and
complexity of the explicit expressions for the perturbed metric
components.

We therefore at the moment do not know  the effect of the
perturbations constructed at the ring singularity and are not as
confident as in the Reissner--Nordstr\"om case that the
perturbations constructed are ``conservative enough'' in their
behavior at the singularity.

In any case, we must stress that the unstable (numerical)
solutions of the Teukolsky equation found here are different from
the algebraically special modes suggested in reference
\cite{cardoso}. In particular, the Starobinsky constant does not
vanish, and, therefore, they represent a new type of solutions of
the Teukolsky equation, of which we have found a few examples,
through some simplifying assumptions, such as taking $m=0$, etc.
It would clearly be interesting to see what happens if these
restrictions are lifted. We are currently working on this problem.

Summarizing, we have shown explicitly in analytic form that the
Reissner--Nordstr\"om space-time is linearly unstable when $Q^2>
M^2$, $M >0$, even in the case the perturbations are
``conservative'' in the sense that they are small at the
singularity in an appropriate sense. We have also numerical
evidence that the Kerr space-time is unstable for $a>M$ at least
for some values of $a,M$. Further  work is needed to confirm that
the instabilities occur for all the range of parameters in
super-extremality.

We wish to thank Vitor Cardoso, Nathalie Deruelle and I\~naki
Olabarrieta for comments on the manuscript. This work was supported in
part by grants of CONICET (Argentina) and the Universidad Nacional de
C\'ordoba. It was also supported in part by grant NSF-PHY-0244335,
NSF-PHY-0554793, NSF-INT-0204937 NASA-NAG5-13430, CCT-LSU and the
Horace Hearne Jr. Institute for Theoretical Physics. G.D. and R.J.G.'s
work is supported by CONICET (Argentina)

\end{document}